\documentclass [10pt]{article}
\usepackage{amsmath}
\usepackage{amsfonts}
\usepackage{amssymb}
\makeatletter \oddsidemargin  0in \evensidemargin 0in
\textwidth16cm \RequirePackage[dvips]{graphicx} \textheight 20.5cm
\newcommand{\singlespacing}{\let\CS=\@currsize\renewcommand{\baselinestretch}{1.5}\tiny\CS}
\makeatletter \oddsidemargin  -.1in \evensidemargin -.1in
\newcommand{\doublespacing}{\let\CS=\@currsize\renewcommand{\baselinestretch}{1.35}\tiny\CS}

\def\@citex[#1]#2{\if@filesw\immediate\write\@auxout{\string\citation{#2}}\fi
  \def\@citea{}\@cite{\@for\@citeb:=#2\do
    {\@citea\def\@citea{,\linebreak[0]\hskip0pt plus .2em}%
      \@ifundefined{b@\@citeb}%
    {{\bf ?}\@warning{Citation `\@citeb' on page \thepage\space undefined}}%
      \hbox{\csname b@\@citeb\endcsname}}}{#1}}

\newtheorem{rule-def}[theorem]{Rule}
\begin{document}
\title{\bf Generalized form of optimal teleportation witnesses}

\author{ Atul Kumar$^{1}$, Satyabrata Adhikari $^{1}$ and ~~Pankaj Agrawal $^{2}$\\
$^{1}$ Centre of Excellence in Systems Science, \\ Indian Institute of Technology Rajasthan, Jodhpur-342011, India\\
$^{2}$ Institute of Physics, Sainik School Post, Bhubaneshwar-751005, Orissa, India \\
}
\date{}
\maketitle{}
\begin{abstract}
We propose a generalized form of optimal teleportation witness to demonstrate their importance in experimental detection of the
 larger set of entangled states useful for teleportation in higher dimensional systems. The interesting properties of our witness
 reveal that teleportation witness can be used to characterize mixed state entanglement using Schmidt numbers. Our results show that
 while every teleportation witness is also a entanglement witness, the converse is not true. Also, we show that a hermitian operator is a
 teleportation witness iff it is a decomposable entanglement witness. In addition, we analyze the practical significance of our study by
 decomposing our teleportation witness in terms of Pauli and Gell-Mann matrices, which are experimentally measurable quantities.
\end{abstract}
PACS numbers: 03.67.-a
\section{Introduction}
Quantum entanglement \cite{einstein,horodecki} is an essential
feature of quantum mechanics which has no classical analogues. The
existence of long range quantum correlations between entangled
particles allows the use of entangled systems as resources for
efficient information transfer through protocols such as quantum
computing \cite{bennett1}, cryptography \cite{gisin},
teleportation \cite{bennett2} and dense coding
\cite{weisner,harrow}. Although entangled states are used for
various theoretical applications in quantum information
processing, the practical use of an entangled resource is
restricted to the successful experimental realization of the
resource. In real experimental set-ups, it is always a challenge
to create and detect entangled states. Also, the prepared
entangled channel may not be robust enough to preserve the
necessary quantum correlations and thus may or may not be
entangled. Hence, successful generation and detection of
entanglement are essential features in any quantum information
processing protocol. For this reason, it is important to develop
efficient and easy to implement experimental set-ups to create and
detect quantum entanglement. Interestingly, class of entanglement
witnesses is necessary and sufficient to detect entanglement
\cite{horodecki1}. The importance of such witness becomes even
more significant due to their decomposition in terms of Pauli spin
matrices (for qubits) and Gell-Mann matrices (for qutrits and
other higher dimensions) which are experimentally measurable
quantities \cite{ganguly}. Recently, a teleportation witness was
proposed to demonstrate whether an underlying quantum state can be
used for teleportation or not \cite{ganguly1}. The measurement of
the expectation value of the witness for unknown states reveals
which states are useful as resource for performing teleportation.
\par In this article, we further address the question of
efficiently distinguishing entangled states for quantum
teleportation using teleportation witness. For this, we propose a
general Hermitian operator and demonstrate that it can be
successfully used as a teleportation witness to differentiate
between entangled classes that can or cannot be used for
teleportation. The interesting properties of our witness show that
it is also an optimal witness of Schmidt number $r~(\geq 2)$ as
compared to the previous work \cite{ganguly1} where the witness is
an optimal witness of Schmidt number 2 only. As the witness
proposed here is optimal, it therefore detects a much larger set
of entangled states in comparison to other witnesses that are not
optimal. In addition, we also prove that all teleportation
witnesses are also entanglement witnesses but every entanglement
witness may not be a teleportation witness. Our results are of
importance not only for practical detection of a larger set of
entangled states useful for quantum teleportation, they also shed
light on the characterization of mixed state entanglement using
Schmidt number. The experimental realization of the witness by
decomposing it into the form of Pauli matrices for $r=2$ and in
the form of Gell-Mann matrices for $r >2$ adds another important
dimension to our study.

\par In section 2, we provide an
introduction to witnesses and then propose our generalized optimal
teleportation witness for the detection of entangled states useful
for the successful completion of teleportation and demonstrate its
properties through theorems and illustrations. Section 3 is
devoted to the application of our teleportation witness in
determining the Schmidt number for the entangled systems. We
discuss the decomposition of teleportation witness in terms of
Pauli and Gell-mann matrices for the experimental detection of
entanglement in section 4. This is followed by the conclusion.

\section{Entanglement and Teleportation Witnesses}
Quantum communication protocols use entanglement between
particles for efficient information transfer from one remote location to another with
reliability. Hence, for an unknown quantum state to be used as a resource in communication protocols, one of the most significant questions we need to answer is whether the underlying quantum state is useful for communication purposes or not. For this, one needs to detect whether the unknown quantum state is entangled or separable. In lower dimensions, Peres-Horodecki criterion provides a necessary and sufficient condition for separability based on the fact that separable states have a positive-partial transpose (PPT) \cite{pereshoro}. However in higher dimensions, the complex nature of quantum entanglement does not allow for a necessary and sufficient condition for its detection. For example, in higher dimensions, one can find entangled states with negative (NPT) as well as positive partial transpose. One solution to this problem is entanglement witnesses \cite{lewenstein,guhne}, which provide a way to detect whether an unknown quantum state is entangled or not. Such witnesses are Hermitian operators with at least one negative eigenvalue; acting as a hyperplane separating entangled states and separable states. The notion of an entanglement witness was extended to Schmidt number witness, which
detects the Schmidt number of quantum states \cite{sperling}. Moreover, witnesses can be separated into two different classes- decomposable and non-decomposable witnesses \cite{woronowicz}. Although non-decomposable witnesses detect both NPT and PPT entangled states, the decomposable witnesses can only detect NPT entangled states. In this article, we provide a necessary and sufficient condition to show that a witness is a teleportation witness if it is decomposable. \par
Quantum teleportation allows for the transmission of arbitrary information from a sender to a receiver using a shared entangled resource between the two. The significant question in this context is the usefulness of the entangled resource for the communication process. We propose a generalized teleportation witness to detect whether an unknown entangled state is useful for teleportation or not. In order to facilitate the discussion of our results, we first briefly describe the fully entangled fraction (FEF) \cite{mhorodecki}, a property of entangled states which is related to the efficacy of quantum teleportation. The fully entangled fraction of a composite system defined by a density matrix $\rho$ can be represented as
\begin{eqnarray}
F(\rho) &=& max_{U}\langle\phi^{+}|U^{\dagger}\otimes
I\rho U\otimes I |\phi^{+}\rangle
\end{eqnarray}
where $|\phi^{+}\rangle=\frac{1}{\sqrt{d}}\sum_{i=0}^{d-1}|ii\rangle$.  The FEF provides a sufficient condition for the determination of states useful for the teleportation process. For example, states in $d \otimes d$ dimensions can always be used as entanglement resources for successful teleportation if the FEF of states is $>$ $1/d$. We propose a Hermitian operator that successfully distinguishes between the states having FEF
$>$ or $\le$ 1/d. Our teleportation witness is optimal and serves as a hyperplane that distinguishes between separable and a larger set of entangled states. We now proceed to propose our generalized  teleportation witness as
\begin{eqnarray}
T_{W} &= &\frac{1}{d}I-|\phi^{+}\rangle\langle\phi^{+}|-\langle
\rho_{0},\frac{1}{d}I-|\phi^{+}\rangle\langle\phi^{+}|\rangle
I \nonumber\\
& = & \left[\langle\phi^{+}|\rho_{0}|\phi^{+}\rangle I-|\phi^{+}\rangle\langle\phi^{+}|\right]
\end{eqnarray}
where $\rho_{0}$ is any reference physical state. In order to show that the Hermitian operator $T_{W}$ is a teleportation witness, we need to prove that the Hermitian operator $T_{W}$ satisfies the following two conditions
\begin{eqnarray}
(i) \langle\sigma, T_{W}\rangle\geq0,~~ \textrm{for~~ all~~
states}~~ \sigma ~~\textrm{which~~ are~~ not~~ useful~~ for~~
teleportation}. \label{cond1}
\end{eqnarray}
\begin{eqnarray}
(ii) \langle \varrho,T_{W}\rangle <0,~~ \textrm{for~~ at~~ least~~
one~~ state}~~ \varrho~~ \textrm{which~~ is~~ useful~~ for~~
teleportation}. \label{cond2}
\end{eqnarray}
where, $\langle\sigma, T_{W}\rangle = {\rm Tr}(\sigma T_{W})$. From Eq. (2), it is clear that our teleportation witness is equivalent to the teleportation witness \cite{ganguly1}
\begin{eqnarray}
W=\frac{1}{d}I-|\phi^{+}\rangle\langle\phi^{+}|
\end{eqnarray}
if $\langle\phi^{+}|\rho_{0}|\phi^{+}\rangle=\frac{1}{d}$. The similarity and differences between the two teleportation witnesses $T_{W}$ and $W$ will be discussed in detail as we move further. For now, we focus on proving that our witness $T_{W}$ is indeed a teleportation witness. For this, we first demonstrate that the operator $T_{W}$ gives a non-negative expectation over all states which are not useful for teleportation. \\

\textbf{Theorem 1:} The hermitian operator
$T_{W}$ is a teleportation witness if
$\langle\phi^{+}|\rho_{0}|\phi^{+}\rangle \geq \frac{1}{d}$.\\
\textbf{Proof:} Let us choose a state $\rho_{0}$ in such a way
that $\langle\phi^{+}|\rho_{0}|\phi^{+}\rangle \geq \frac{1}{d}$. The operator $T_{W}$ would be a teleportation witness if it satisfies the conditions (\ref{cond1}) and (\ref{cond2}).\\
(i) Let $\sigma$ be an arbitrary state chosen from the set which is not useful for
teleportation, i.e. $F(\sigma)\leq\frac{1}{d}$. Hence, we have
\begin{eqnarray}
\langle \sigma,T_{W}\rangle &&= \langle \sigma,
\left[\langle\phi^{+}|\rho_{0}|\phi^{+}\rangle I-|\phi^{+}\rangle\langle\phi^{+}|\right]\rangle \nonumber\\&& =
\left[\langle\phi^{+}|\rho_{0}|\phi^{+}\rangle - \langle\phi^{+}|\sigma|\phi^{+}\rangle \right] \nonumber\\&&
\geq
\left[\langle\phi^{+}|\rho_{0}|\phi^{+}\rangle-max_{U}\langle\phi^{+}|U^{\dagger}\otimes
I\sigma U\otimes I |\phi^{+}\rangle\right] \nonumber\\&&
=\left[\langle\phi^{+}|\rho_{0}|\phi^{+}\rangle - F(\sigma)\right] \nonumber\\&&
\geq0
\end{eqnarray}
Therefore, the expectation value of the Hermitian operator $T_{W}$ is non-negative for all bipartite $d$-dimensional states which are not useful for teleportation. Hence, the Hermitian operator $T_{W}$ satisfies (\ref{cond1}).\\
(ii) To prove that our witness detects at least one entangled state $\varrho$ that is useful for teleportation, we fix $\rho_{0}$ and $\varrho$ as
\begin{eqnarray}
\rho_{0}=
k|\phi^{+}\rangle\langle\phi^{+}|+\frac{1-k}{d^{2}}I,~~~0\leq
k\leq1
\end{eqnarray}
and
\begin{eqnarray}
\varrho=
\beta|\phi^{+}\rangle\langle\phi^{+}|+\frac{1-\beta}{d^{2}}I,
~~~\frac{-1}{d^{2}-1}\leq \beta\leq1
\end{eqnarray}
where $\varrho$ is an isotropic state and is entangled $\forall \, \,  \beta  > \frac{1}{{d + 1}}$. Moreover, the condition $\langle\phi^{+}|\rho_{0}|\phi^{+}\rangle \geq \frac{1}{d}$ shows that
\begin{eqnarray}
k \geq \frac{1}{d+1}
\end{eqnarray}
Hence, $\langle \varrho,T_{w}\rangle$ can be rewritten as
\begin{eqnarray}
\langle \varrho,T_{W}\rangle &&= \langle \varrho,
\left[\langle\phi^{+}|\rho_{0}|\phi^{+}\rangle I-|\phi^{+}\rangle\langle\phi^{+}|\right]\rangle\nonumber\\&&
=[\langle\phi^{+}|\rho_{0}|\phi^{+}\rangle)-\langle\phi^{+}|\varrho|\phi^{+}\rangle]
{}\nonumber\\&&=\frac{(k-\beta)(d^{2}-1)}{d^{2}}<0,~~~{\rm when}~~
k<\beta
\end{eqnarray}
Therefore, our witness $T_{W}$ detects at least one state useful for
teleportation when $\frac{1}{d+1} \leq k<\beta$ and thus satisfies (\ref{cond2}). Our analysis shows that all entangled isotropic states can be successfully used for teleportation, which is a well known result \cite{zhao}. This completes the proof that our witness satisfies both the required conditions (\ref{cond1}) and (\ref{cond2}) and thus is a teleportation witness. \par
The form of teleportation witness $T_{W}$ suggests that one can also replace $\langle\phi^{+}|\rho_{0}|\phi^{+}\rangle = s$; a real constant, but for reasons to be described below, we would like to retain the original form of our operator $T_{W}$ = $\langle\phi^{+}|\rho_{0}|\phi^{+}\rangle I-|\phi^{+}\rangle\langle\phi^{+}|$. For example, if one uses the form $T'_{W}$ = $s I-|\phi^{+}\rangle\langle\phi^{+}|$, then the condition $\langle \varrho,T'_{W}\rangle < 0$ requires that $\beta > s$. Hence, the operator detects states useful for teleportation when $\frac{1}{d} \leq s<\beta$ which shows that the operator $T'_{W}$ detects a smaller set of entangled states with respect to $T_{W}$, at least in this particular case. \\

\textbf{Theorem 2:} Every teleportation witness in $d \otimes d$ dimensions $(d > 2)$ is an entanglement witness, but every entanglement witness in $d \otimes d$ dimensions $(d > 2)$ may not be a teleportation witness.\\
\textbf{Proof:} A Hermitian operator $A$ is an entanglement witness if it satisfies the following two conditions: \\
(i) The expectation value of the operator $A$ must always be non-negative for all separable states in $d \otimes d$ dimensions i.e. for states with FEF $\le$ $\frac{1}{d}$. \\
(ii) The expectation value of the operator $A$ must be negative for at least one entangled state in $d \otimes d$ dimensions
i.e. for a state with FEF $>$ $\frac{1}{d}$. \\
Teleportation witnesses also satisfy above two conditions as required by the conditions (\ref{cond1}) and (\ref{cond2}). Hence every teleportation witness is also an entanglement witness.  \\
Conversely, we can always show that there exists an indecomposable entanglement witness which would not satisfy the condition (\ref{cond2}) and therefore cannot be termed as a teleportation witness. For example, one can always find at least one bound entangled state for which the expectation value of non-decomposable entanglement witnesses would be negative \cite{chruscinski}. Thus, non-decomposable entanglement witnesses detect at least one bound entangled state which is not useful for teleportation (since the singlet fraction of a bound entangled state is equal to $\frac{1}{d}$). Hence, non-decomposable entanglement witnesses do not satisfy the required condition (\ref{cond2}) to be a teleportation witness. This completes the proof that in $d \otimes d$ dimensions $(d > 2)$, every teleportation witness is also an entanglement witness, however the converse is not true. \\

\textbf{Corollary 1:} In $2 \otimes 2$ dimensions, every teleportation witness is an entanglement witness and {\em vice versa}. \\
\textbf{Proof:} In $2 \otimes 2$ dimensions, every entangled bi-partite state can be made useful for teleportation up to stochastic local operation and classical communication (SLOCC) \cite{verstraete}. Thus teleportation and entanglement witnesses will always satisfy the conditions (\ref{cond1}) and (\ref{cond2}). Hence, every teleportation witness in $2 \otimes 2$ dimensions will also be an entanglement witness and {\em vice versa}.  \\

\textbf{Corollary 2:} A witness is a teleportation witness iff it is decomposable.\\
\textbf{Proof:} By definition, the expectation value of teleportation witnesses for all bound entangled states is always
non-negative. Using theorem 2, we have shown  that every teleportation witness is an entanglement witness and since teleportation witnesses cannot detect bound entangled states, a teleportation witness can only be a decomposable entanglement witness. \\
Conversely, a decomposable entanglement witness can only detect NPT states i.e. states for which the expectation value of the decomposable entanglement witness would be negative {(\ref{cond2})}. As it also satisfies the condition (\ref{cond1}), a decomposable entanglement witness is also a teleportation witness.

\section{Optimal Teleportation Schmidt Witness}
In this section, we show that our generalized teleportation witness is also an optimal witness. In addition, we analyze the Schmidt number of an arbitrary mixed state useful for teleportation by constructing a Schmidt number teleportation witness. The results obtained in this section provide a way to characterize the mixed state entanglement in bi-partite systems. For example, we propose a form of teleportation Schmidt witness to demonstrate its significance in calculating the Schmidt number of mixed entangled states for a given range of parameters. For bipartite systems, Schmidt number indicates the number of degrees of freedom that are entangled between two subsystems.   \\

\textbf{Theorem 3:} Teleportation witness $T_{W}$ with $\langle\phi^{+}|\rho_{0}|\phi^{+}\rangle \geq \frac{1}{d}$ in $d \otimes d$ dimensions is an optimal teleportation witness of Schmidt number $r \geq 2$.\\
\textbf{Proof:} An optimal entanglement witness of Schmidt number r in $d \otimes d$ dimensions is given by \cite{sanpera}
\begin{eqnarray}
W_{opt}=I-\frac{d}{r-1}|\phi^{+}\rangle\langle\phi^{+}|
\label{optwit}
\end{eqnarray}
If $r=2$, we have
\begin{eqnarray}
W_{opt}&=&I-d|\phi^{+}\rangle\langle\phi^{+}|{}\nonumber\\&
=&d(\frac{1}{d}I-|\phi^{+}\rangle\langle\phi^{+}|){}\nonumber\\&
\propto & W ~~~{\rm from [Eq. (5)]}~~ \nonumber\\&
= & T_{W} ~~~{\rm if}~~ \langle\phi^{+}|\rho_{0}|\phi^{+}\rangle = \frac{1}{d}
\end{eqnarray}
Hence, teleportation witnesses $W$ and $T_{W}$ are optimal teleportation witnesses of Schmidt number 2. For $\langle\phi^{+}|\rho_{0}|\phi^{+}\rangle > \frac{1}{d}$, the teleportation witness $T_{W}$ can be re-expressed as
\begin{eqnarray}
T_{W} &=&\langle\phi^{+}|\rho_{0}|\phi^{+}\rangle I-|\phi^{+}\rangle\langle\phi^{+}| \nonumber\\
&=&\langle\phi^{+}|\rho_{0}|\phi^{+}\rangle
\left[I-\frac{1}{\langle\phi^{+}|\rho_{0}|\phi^{+}\rangle}|\phi^{+}\rangle\langle\phi^{+}|\right]
\label{telwit}
\end{eqnarray}
Comparing Eqs. (\ref{optwit}) and (\ref{telwit}) shows that the
teleportation witness $T_{W}$ is proportional to the optimal
witness $W_{opt}$ if
\begin{eqnarray}
r &= & d\langle\phi^{+}|\rho_{0}|\phi^{+}\rangle+1 {}\nonumber\\&>&2,
~~{\rm since}~~\langle\phi^{+}|\rho_{0}|\phi^{+}\rangle>\frac{1}{d}
\end{eqnarray}
This completes the proof that our teleportation witness $T_{W}$ is an optimal witness of Schmidt number $r \geq 2$.\\

\textbf{Illustration:} Let us consider a family of states for $d\times d$ dimensional systems
\begin{eqnarray}
\chi_{\beta}=\beta|\phi^{+}\rangle\langle\phi^{+}|+\frac{1-\beta}{d^{2}}I,
~~\frac{-1}{d^2-1}\leq\beta\leq1
\end{eqnarray}
where the state $\chi_{\beta}$ is entangled $\forall \, \,
\beta>\frac{1}{d+1}$. In this illustration, we propose the form of our teleportation Schmidt
witness to demonstrate that it identifies the ranges of parameters for which
the state in Eq. (15) is of Schmidt number $r \geq 2$.\\
For this, we fix the reference state $\rho_{0}$ as
\begin{eqnarray}
\rho_{0}=\frac{1-f_{0}}{d^{2}-1}I+\frac{d^2f_{0}-1}{d^{2}-1}|\phi^{+}\rangle\langle\phi^{+}|
\end{eqnarray}
which gives $f_{0}=\langle\phi^{+}|\rho_{0}|\phi^{+}\rangle$, satisfying
$\frac{1}{d}\leq f_{0}\leq1$.\\
Also, the singlet fraction of $\rho_{0}$ is given by \cite{zhao}
\begin{eqnarray}
F(\rho_{0})= f_{0}, ~~~~
\frac{1}{d}\leq f_{0}\leq 1
\end{eqnarray}
Thus, the expectation value of the teleportation witness $T_{W}$ in the state $\chi_{\beta}$ is
\begin{eqnarray}
Tr(T_{W}\chi_{\beta})=\frac{(d^2f_{0}-1)-\beta(d^{2}-1)}{d^{2}}
\end{eqnarray}
If we take $f_{0}=\frac{1}{d}$ then the teleportation witness $T_{W}$ assures that the state $\chi_{\beta}$, useful for
teleportation, is of Schmidt number $2$ when $\beta \in \left(\frac{1}{d+1},\frac{2d-1}{d^{2}-1}\right]$. In general, if we take
$f_{0}=\frac{r-1}{d}, (r=2,3,....d)$ then the corresponding teleportation witness $(T_{W})_{r}$ assures that the state
$\chi_{\beta}$, useful for teleportation, is of Schmidt number $r$ when $\beta \in \left(\frac{d(r-1)-1}{d^2-1},\frac{dr-1}{d^{2}-1}\right]$. For example, in $3 \otimes 3$ dimensions the state $\chi_{\beta}$ is entangled for $\forall \, \, \beta>\frac{1}{4}$ and the range of $\beta$ confirms that $2 < r \leq 3$. Therefore, the Schmidt number of the state $\chi_{\beta}$ must be 3 for the given range of $\beta$. Similarly, one can calculate the Schmidt number for higher dimensional systems for the given range of parameters. Hence, our teleportation Schmidt witness can also be used to characterize mixed state entanglement in terms of Schmidt numbers.

\section{Experimental determination of $T_{W}$}
In this section, we analyze the decomposition of our teleportation witness in terms of Pauli spin matrices for $2 \otimes 2$ systems, and Gell-Mann matrices for higher dimensional systems. Such decompositions allow for the experimental detection of entanglement in an unknown state through the measurements of expectation values of teleportation witness. For experimental realization of teleportation witnesses, we need to decompose them into projectors of the form \cite{guhne}
\begin{eqnarray}
{T_W} = \sum\limits_{i = 1}^j {{k_i}\left| {{e_i}} \right\rangle } \left\langle {{e_i}} \right| \otimes \left| {{f_i}} \right\rangle \left\langle {{f_i}} \right|
\end{eqnarray}
If $\langle\phi^{+}|\rho_{0}|\phi^{+}\rangle=\frac{1}{2}$ then the form of $T_{W}$ is equivalent to the teleportation witness $W$ in Eq. (5) and thus can be decomposed in form of Pauli spin matrices such that
\begin{eqnarray}
{T_W} = \frac{1}{2}[I \otimes I - {\sigma _x} \otimes {\sigma _x} + {\sigma _y} \otimes {\sigma _y} - {\sigma _z} \otimes {\sigma _z}]
\end{eqnarray}
Eq. (20) indicates that the total number of measurements required
to estimate our witness is limited to 3 as compared to the
measurement of 15 parameters required for full state tomography.
This difference in number of measurements is even more evident for
higher dimensional systems providing a practical utility to our
results when compared to full state tomography in distinguishing
the useful states for quantum teleportation. Similarly, for $d
\otimes d$ systems our teleportation witness can be decomposed as
\begin{eqnarray}
T_{W}^{d} = \left( {\left\langle {{\phi ^ + }} \right|{\rho _0}\left| {{\phi ^ + }} \right\rangle  - \frac{1}{{{d^2}}}} \right)I - \frac{1}{{2d}}\Lambda
\end{eqnarray}
where
\begin{eqnarray}
\Lambda  = \sum\limits_{i < j} {\Lambda _s^{ij} \otimes } \Lambda _s^{ij} - \sum\limits_{i < j} {\Lambda _a^{ij} \otimes } \Lambda _a^{ij} + \sum\limits_{m = 1}^{d-1} {{\Lambda ^m}}  \otimes {\Lambda ^m}
\end{eqnarray}
The suffix s, a and superscript m represent the symmetric, antisymmetric and diagonal Gell-Mann matrices \cite{bertlmann}, respectively. For example, in case of qutrits i.e. $3 \otimes 3$ dimensional systems the teleportation witness $T_{W}$ can be represented as
\begin{eqnarray}
T_{W}^{3} = \left( {\left\langle {{\phi ^ + }} \right|{\rho _0}\left| {{\phi ^ + }} \right\rangle  - \frac{1}{{{9}}}} \right)I - \frac{1}{{6}}\Lambda
\end{eqnarray}
where
\begin{eqnarray}
\Lambda  &=& \Lambda _s^{12} \otimes \Lambda _s^{12} + \Lambda _s^{13} \otimes \Lambda _s^{13} + \Lambda _s^{23} \otimes \Lambda _s^{23} - \Lambda _a^{12} \otimes \Lambda _a^{12} - \Lambda _a^{13} \otimes \Lambda _a^{13} - \Lambda _a^{23} \otimes \Lambda _a^{23} \nonumber \\
&+& {\Lambda ^1} \otimes {\Lambda ^1}{\rm{  + }}{\Lambda ^2} \otimes {\Lambda ^2}{\rm{ }}
\end{eqnarray}
and
$\Lambda _s^{12} = \left( {\begin{array}{*{20}{c}}
0&1&0\\
1&0&0\\
0&0&0
\end{array}} \right)$, $\Lambda _s^{13} = \left( {\begin{array}{*{20}{c}}
0&0&1\\
0&0&0\\
1&0&0
\end{array}} \right)$, $\Lambda _s^{23} = \left( {\begin{array}{*{20}{c}}
0&0&0\\
0&0&1\\
0&1&0
\end{array}} \right)$, $\Lambda _a^{12} = \left( {\begin{array}{*{20}{c}}
0&-i&0\\
i&0&0\\
0&0&0
\end{array}} \right)$,

$\Lambda _a^{13} = \left( {\begin{array}{*{20}{c}}
0&0&-i\\
0&0&0\\
i&0&0
\end{array}} \right)$, $\Lambda _a^{23} = \left( {\begin{array}{*{20}{c}}
0&0&0\\
0&0&-i\\
0&i&0
\end{array}} \right)$, $\Lambda ^{1} = \left( {\begin{array}{*{20}{c}}
1&0&0\\
0&-1&0\\
0&0&0
\end{array}} \right)$, $\Lambda ^{2} = \frac{1}{\sqrt{3}}\left( {\begin{array}{*{20}{c}}
1&0&0\\
0&1&0\\
0&0&-2
\end{array}} \right).$ \\
 Hence, the decomposition of our teleportation witness in terms of measurable quantities not only allows for the experimental detection of entanglement but also provides a way to decrease the requirement of number of measurements when compared to full state tomography.

\section{Conclusion}
We have proposed an optimal teleportation witness and demonstrated its practical utility in distinguishing the entangled states useful for teleportation. The form of teleportation witness $T_{W}$ studied here is a generalized form of the witness $W$ discussed in \cite{ganguly1}. The results obtained in this article are different from the previous study \cite{ganguly1} in a sense that the witness $W$ is an optimal witness of Schmidt number 2, but our witness $T_{W}$ is an optimal witness of Schmidt number $\geq 2$. This allows for the detection of larger sets of mixed entangled states useful for teleportation in higher dimensions as well. We found that all the teleportation witnesses are also entanglement witnesses, however the converse is not true. It also turned out that a teleportation witness is always a decomposable entanglement witness, which is a necessary and sufficient condition. The experimental determination of such teleportation witnesses decreases the number of parameters to be measured in comparison to the full state tomography of an unknown state useful for teleportation, indicating the practical significance of our study. In future, we would like to study the utility of such witnesses for other communication protocols as well. Another question of particular interest would be the form and characteristics of teleportation witnesses in multiqubit systems.

\end{document}